# The Dynamic Shift of Neuron Excitability Observed with Enveloped High Frequency Stimulation

Jiahui Wang[a,b,c], Hao Wang[a,b,c,1], Xin Yuan Thow[b], Nitish V. Thakor[a,b], Chengkuo Lee[a,b,c,1]

[a]Department of Electrical & Computer Engineering, National University of Singapore, 4 Engineering Drive 3, 117576, Singapore; [b]Singapore Institute for Neurotechnology (SINAPSE), National University of Singapore, 28 Medical Drive, #05-COR, 117456, Singapore; [c]Center for Intelligent Sensors and MEMS, National University of Singapore, 4 Engineering Drive 3, 117576, Singapore

[1]To whom correspondence may be addressed. Email: elewangh@nus.edu.sg or elelc@nus.edu.sg.

**Abstract**

Electrical stimulation is an artificial way to initiate action potentials in excitable neural tissues. We conducted pure physical reasoning of electrical stimulation influence on transmembrane ion distribution and found that neuron excitability can be affected to either increase (stronger force) or decrease (fatigue). Although it has been widely observed that electrical stimulation results in muscle fatigue, there is no previous report on increased force-generating capability of muscle during electrical stimulation. We designed enveloped high frequency stimulation (EHFS), which was applied on muscle tissue in an acute rat model. We discovered that EHFS dynamically shifts neuron excitability, which manifests itself as either increase or decrease of muscle force-generating capability. To find out why such dynamic neuron excitability shift could be observed with EHFS, we modeled using the static distributed-parameter circuit model as proposed in our previous work, and then confirmed by *in vivo* measurements. It turns out that EHFS achieves synchronized recruitment of motoneurons at the two stimulation electrodes connected to the positive and negative terminal of a stimulator. Then, with the considerations of neuron excitability shift, a dynamic model is built to study the generation and propagation of action potentials in external electrical stimulation, which explains the widely-observed phenomena of collision blocking in peripheral nerve stimulation. Thus, our observations on neuron excitability shift offer a glimpse into how external electrical stimulation dynamically interact with neural tissues, and guide novel electrical stimulation strategies to activate (or block) neural signals.

**Significance**

Electrical stimulation is thought to be an artificial way to initiate action potentials in excitable neural tissues. Here, we discovered that electrical stimulation dynamic shifts neuron excitability, which manifests itself as either increase or decrease of muscle force-generating capability, by applying specially designed enveloped high frequency stimulation on muscle tissue in an acute rat model. Our results provide conclusive *in vivo* evidence to confirm dynamic neuron excitability shift caused by external electrical stimulation. It raises the possibility that the dynamic interaction between electrical stimulation and neural tissue is a previously unappreciated mechanism to account for important phenomena during electrical stimulation, including muscle fatigue, and collision blocking.

Functional electrical stimulation (FES) aims at recovering meaningful functions, such as walking and grasping, to limbs paralyzed by nerve damage through the application of electrical stimulation on their nerves or muscles. However, compared to voluntarily controlled movements, electrical stimulation rapidly causes fatigue, which is characterized as decrease of force-generating capability of muscle. The reason for increased fatigability using FES has been thought to be twofold. On the one hand, FES preferably recruits large-diameter motoneurons at lower input current amplitude, which is opposite to the normal physiological order (1-6). On the other hand, FES tends to recruit all motoneurons simultaneously (7-11).

Apart from the above well-established explanations for fatigue in FES, we speculate that FES might induce disturbance of transmembrane ion ($Na^+$ and $K^+$) distribution. For voluntarily controlled movements, disturbance of transmembrane ion distribution has been found to reduce ion channel excitability, which leads to force output reduction during intense fatiguing exercises (12, 13). *In vitro* observations confirmed that both decreased extracellular $Na^+$ concentration (14, 15) and increased extracellular $K^+$ concentration (16-18) reduce ion channel excitability. In this paper, we will study whether external electrical stimulation distorts transmembrane ion distribution, and how this mechanism is related to increased fatigability in FES.

During neuromuscular stimulation, motoneurons are excited to control muscle fibers (19). A successful excitation of motoneurons starts with the generation of action potentials. In voluntarily controlled movements, one action potential will sequentially trigger the opening of ion channels on the next node of Ranvier, so that this action potential can be relayed along the motoneuron axon. However, when it comes to electrical stimulation, the transmembrane voltage resulting from electrode-tissue interaction determines the opening of ion channels. Unlike natural action potentials, this transmembrane voltage not only initiates the opening of ion channels, but also distorts ion distribution by lingering on after the ion channels open.

To understand electrical stimulation influence on transmembrane ion distribution, Fig. 1 shows a general illustration of pure physical reasoning on how external electrical stimulation distorts ion distribution. Existence of positive external stimulation results in a larger $Na^+$ influx and smaller $K^+$ efflux, which reduces the threshold voltage for the following electrical stimulation. Similarly, existence of negative external stimulation increases the threshold voltage, making ion channels more difficult to open in the following stimulation. Thus, electrical stimulation has bidirectional effects on force output, by either increasing or decreasing ion channel excitability. However, it has only been reported that electrical stimulation induces muscle fatigue (decreased neuron excitability) (1-11), and no previous observations on increased muscle force-generating capability during electrical stimulation have been reported.

In this paper, we discovered that electrical stimulation on muscle tissue induces increase or decrease of neuron excitability, by applying specially designed enveloped high-frequency stimulation (EHFS) on an acute rat model (Fig. 2A). With the probability calculation for neuron excitation (20) and the static distributed-parameter circuit model for muscle tissue (21) proposed in our previous work, we modeled the resistance of neuron excitability to distorted ion distribution at different input current amplitude, and then confirmed with *in vivo* measurements that large increase and decrease of force output tend to occur at low current stimulation. Unlike the conventionally used square wave input current which induces voltage waveforms of reversed polarity at two stimulation electrodes connected to the positive and negative terminal of a stimulator (21), EHFS achieves almost the same waveform that induces synchronized neuron recruitment at both electrodes, which was confirmed by modeling and *in vivo* measurement. Thus, EHFS induces neuron excitability shift of the same direction at both electrodes (to decrease or increase together), to enable the observations of large increase and decrease of force output.

Since we have found that external electrical stimulation shifts neuron excitability by distorting transmembrane ion distribution, action potentials formed by transmembrane ion flows will also shift local neuron excitability. We further built a dynamic model to study the generation and propagation of action potentials in electrical stimulation on the peripheral nerves. This dynamic model helps to elucidate important phenomena, such as collision blocking.

**Results**
**Force profiles generated by enveloped high frequency stimulation (EHFS).** Two stainless steel wires were sutured along the muscle fiber direction, with a separation of around 1 cm and a depth into the muscle of 2-3 mm (Fig. 2A), to deliver specially designed enveloped high frequency stimulation (EHFS in Fig. 2B) to the Tibialis Anterior (TA) muscle in an acute rat model. Every 1 s, a train of 10 envelops were delivered. Among the 10 envelops, one envelop was delivered every 16.7 ms. Each envelop consisted of multiple high frequency inner sinewave current pulses (20 kHz). The amplitude of these inner sinewave pulses was modulated by a relatively low frequency outer sinewave (each envelop lasted for 1 ms). In this way, the EHFS had both gradual onset and ending. When the TA muscle was stimulated, leg would kick forward, and force was measured by a force gauge tied to the ankle.

Fig. 2C1-C3 show three independent force profiles measured from three subjects stimulated with EHFS, which was continuously delivered throughout the stimulation trial in the way described in Fig. 2B. These three force profiles show force decrease (decrease to even zero force output) followed by force increase (increase to even higher than the initial force at the beginning of stimulation). Here, the initial force, the first zero force output, and the first recovery period are referred to as area A, B, and C, respectively. These are very different from the conventionally reported force profiles generated by square wave stimulation (4, 22-24), which usually show a monotonically decreasing trend after the initial force reaches the maximum.

To interpret our measured force profiles, we need to refer back to the physical reasoning on how external electrical stimulation distorts an action potential as discussed in the introduction (Fig. 1). Two generalized scenarios are expected: external electrical stimulation may facilitate the influx of $Na^+$ to decrease stimulation threshold for the following stimulation, and external electrical stimulation may also facilitate the efflux of $K^+$ to increase stimulation threshold. This explains what we have observed. The decrease of force output from area A to area B is in correspondence with decreased neuron excitability caused by larger $K^+$ efflux. The neuron excitability decreases to a point when external stimulation cannot activate motoneurons anymore, and it results in the zero output of area B. Since external stimulation is still continuously delivered to muscle, some $Na^+$ channels open although the generated $Na^+$ flow is not large enough to trigger an action potential. This $Na^+$ inflow helps to recover neuron excitability until a point, when external stimulation can excite motoneurons again in area C. Then, motoneurons will go through multiple rounds of excitability changes until energy is finally depleted and no force output can be further generated.

Thus, up to this point, the pure physical reasoning in Fig. 1 can well explain our measured force profiles, which indicates that external electrical stimulation dynamically shifts neuron excitability. Then, the question arises, why this can be observed with EHFS. Our investigation includes three steps: Firstly, with modeling, we compared input current of square wave, high frequency sinewave, and EHFS, in terms of the voltage waveforms generated at the two stimulation electrodes (connected to the positive and negative terminal of a stimulator); Secondly, we studied the resistance of neuron excitability to distorted transmembrane ion distribution at different input current amplitude, by modeling and then *in vivo* measurements; Thirdly, we modeled the synchronization of voltage waveforms generated at the two stimulation electrodes with EHFS at different frequency and current amplitude, and confirmed with *in vivo* measurements.

**Comparison of voltage waveforms in response to different input current waveforms with modeling.** As proved in our previous paper (21), when square wave input current is applied, voltage waveforms generated at the two stimulation electrodes are of the opposite polarity, which results in a different pattern of exceeding the threshold voltage at the two electrodes. It means that the recruitment of motoneurons at the two electrodes is different. Voltage waveforms generated at the two stimulation electrodes using square wave, high frequency sinewave, and EHFS were modeled and compared.

A distributed-parameter circuit model (Fig. 3A) was used to model voltage waveforms generated at the two stimulation electrodes (motoneuron P1 next to the positive electrode E+, and motoneuron P2 next to the negative electrode E-). In this distributed-parameter circuit model, extracellular medium is modeled as resistor, myelin is modeled as inductor, and cell membrane is modeled as capacitor. In this way, RLC components (resistor, inductor, capacitor) represent motoneurons, and RC components (resistor, inductor, capacitor) represent muscle fibers. We firstly modeled the voltage waveform using this distributed-parameter circuit (21), and then calculated the probability of neuron excitation using the method proposed in our previous work (20). For each input current amplitude, the probability of neuron excitation was calculated using the effective voltage waveform which exceeded voltage threshold.

Fig. 3B shows the voltage waveform on P1, which is synchronized with the input EHFS current. Fig. 3C, Fig. 3D, and Fig. 3E show the voltage waveforms and probability curves (probability of neuron excitation with respect to input current amplitude) of P1 and P2, generated by square wave (positive-first biphasic waveform, with 500 µs duration for each phase), high frequency sinewave (20 kHz and last for 1 ms), and EHFS (20 kHz inner wave frequency, 0.5 kHz outer envelope frequency, with 1 ms duration), respectively. For square wave current input, voltage waveforms on P1 and P2 have opposite polarity and exceed threshold voltage in different patterns, and thus the probability curves show a wide discrepancy for P1 and P2. For high frequency sinewave current input, the synchronization of voltage waveforms on P1 and P2 is much improved compared to square wave stimulation, because of the fast alternation of input current polarity. However, at the onset and ending of the high frequency sinewave stimulation, voltage waveforms on P1 and P2 still show asynchrony, which leads to the discrepancy of the probability curves for P1 and P2. When it comes to EHFS, voltage waveforms on P1 and P2 are highly synchronized, so that the probability curves are overlapped. Unlike the other input current waveforms, voltage waveforms generated by the EHFS closely follow the pattern of input current waveform without showing any peak and tail at the onset and ending of stimulation.

Thus, EHFS achieves synchronized recruitment of motoneurons at the two stimulation electrodes, while square wave current fails to achieve this. This synchronization of motoneuron recruitment is crucial to generate an output force profile with large decrease and increase, because it will induce neuron excitability shift in the same direction at the two stimulation electrodes (to increase or to decrease together). In the case of asynchronous motoneuron recruitment, neuron excitability shift varies for different motoneurons. We also observed decrease and increase of force output caused by neuron excitability shift under continuous square wave current input (Fig. S1). However, the change of force profile can be easily observed when EHFS is used. In other words, EHFS amplifies neuron excitability shift, due to the synchronized neuron recruitment at the two stimulation electrodes.

However, not all the EHFS can achieve an output force profile with large decrease and increase. There are two important parameters of the EHFS: the amplitude of the outer sinewave, and the frequency of the inner sinewave. We will investigate how these two parameters affect output force profiles in the following sessions.

**Resistance of neuron excitability to distorted transmembrane ion distribution is affected by input current amplitude.** There are two observations related to small input current amplitude: Firstly, it generates small force; Secondly, small changes in input current at small amplitude tend to result in

unstable force output (Fig. S2, Fig. S3). Thus, the question arises, whether these two observations are correlated. To address this question, we modeled the resistance of neuron excitability to distorted transmembrane ion distribution, using the distributed-parameter circuit model in Fig. 4A. Then, we used the quantized *in vivo* measurements to confirm the modeling results.

As discussed in the earlier session (Fig. 1), external electrical stimulation distorts transmembrane ion distribution during an action potential. To study input current amplitude influence on output force stability, we need to model how input current amplitude affects neuron excitation probability. Input current of different amplitude distorts transmembrane ion distribution to different extent, which results in different resting potential of ion channels. Here, we use threshold voltage shift to characterize the change of resting potential. With this threshold voltage shift, we can quantitatively model the input current amplitude influence on neuron excitation probability. Fig. 4B shows a modeling demonstrating this. At different current amplitude, the same $V_{Threshold}$ (threshold voltage) and $\Delta V_{Threshold}$ (threshold voltage shift) is applied. Due to $\Delta V_{Threshold}$, the neuron excitation probability is changed, which is characterized by $\Delta Probability$ (change of neuron excitation probability). In Fig. 4C, $\Delta Probability$ increases and then decreases with increasing input current amplitude, at all frequencies. This is because large input current amplitude generates large force, which is more resistive to a same force change as compared to small force.

*In vivo* measurements confirmed that larger current amplitude generates more stable output force. Fig. 4D shows the quantized force unstabilization (defined as the ratio of standard deviation of force and maximum force) of four force profiles measured at different input current amplitude in Fig. 4E. In Fig. 4C, during a narrow range at low current amplitude, $\Delta Probability$ increases with increasing current, which represents the transition from stable zero output to non-zero output when motoneuron just starts to be excited. This current range is very narrow, and we did not catch a current amplitude that falls in this range during *in vivo* measurements. However, the measurement curve matches the modeling curve outside that small amplitude range, as force becomes more stabilized at large current amplitude.

Therefore, EHFS at small current amplitude reveals unstable output force profile with large decrease and increase. However, there is another crucial uniqueness about this EHFS, which enables the clear observation of motoneuron excitability shift. How frequency affects the synchronization of neuron recruitment at the two stimulation electrodes will be discussed in the following session.

**The synchronization of neuron recruitment is affected by frequency and amplitude of enveloped high frequency stimulation (EHFS).** Beyond just the unique envelope profile, two different EHFS parameters were studied to investigate their effect on the synchronized neuron recruitment, namely, the frequency of the inner sinewave and the amplitude of the outer sinewave.

When positive-first EHFS is delivered to muscle, neurons at the positive electrode experience positive-first voltage waveform, while neurons at the negative electrode experience negative-first voltage waveform. To know the difference of neuron recruitment at the two stimulation electrodes, we need to study how neuron responds differently to positive-first and negative-first voltage waveform. Here, the modeling was performed using the distributed-parameter circuit model in Fig.4A.

In the modeling, positive-first and negative-first EHFS of different inner sinewave frequency (from 1 kHz to 5 kHz, with 0.2 kHz step) and outer sinewave current amplitude (from 0.05 mA to 5 mA, with 0.05 mA step) were used as input. Fig. 5A shows EHFS with increasing frequency from 1 kHz to 5 kHz, where we can visually tell that positive-first and negative-first EHFS become more synchronized at higher frequency. The neuron excitation probability curves in response to positive-first and negative-first EHFS are shown in Fig. 5B and Fig. 5C, respectively. The difference between neuron excitation probabilities (defined as probability difference here) in response to positive-first and negative-first EHFS was calculated and shown in Fig. 5D. In Fig. 5D, increasing frequency affects probability difference in two ways: Firstly, the amplitude of probability difference reduces; Secondly, probability difference curves experience a polarity shift. It means that the stimulation efficiency of positive-first and negative-first EHFS reverse within a certain frequency range. Apart from the frequency of inner sinewave, the current amplitude of the outer sinewave also plays an important role in modulating probability difference. At large current amplitude, probability difference reduces towards zero at all frequencies. It means that probability difference can only be clearly observed at relatively small current amplitude.

We then confirmed the influence of increasing frequency and amplitude on synchronization of neuron recruitment with *in vivo* measurements. To avoid accumulated muscle fatigue, trains (a train consists of 10 envelops) of positive-first and negative-first EHFS were alternatively applied every 1 s (Fig. 6A). If neuron recruitment at the two stimulation electrodes is synchronized, a smooth force profile without abrupt change in between each generated force output is expected. Otherwise, the output force at even position is expected to be different from odd position. Fig. 6B shows the output force difference generated by positive-first and negative-first EHFS, which matches well with the three features of modeling results in Fig. 5D. Firstly, force difference decreases with increasing frequency and is largest at low frequency of 1 kHz. Secondly, the polarity of force difference reverses from positive to negative at 2 kHz, and then reverses back from negative to positive at 4 kHz. Thirdly, force difference decreases with increasing current amplitude, and is only observable at relatively small current amplitude. The detailed force profiles are shown in Fig. 6C, which clearly shows the output force difference at low frequency and low current amplitude.

**Discussion**

**External electrical stimulation shifts neuron excitability.** In this paper, enveloped high-frequency stimulation (EHFS) on muscle tissue revealed that external electrical stimulation shifts neuron excitability. We confirmed how the two important parameters of EHFS (amplitude of outer sinewave, and frequency of inner sinewave) enabled the observations of increase and decrease of force output. One the one hand, small amplitude of outer sinewave results in low force output, which is more vulnerable to neuron excitability shift. On the other hand, proper current amplitude of outer sinewave together with high frequency of inner sinewave contributes to synchronized neuron recruitment at the two stimulation electrodes connected to the positive and negative terminal of a stimulator.

Up to this point, we have already known that external electrical stimulation dynamically shifts neuron excitability. This helps to understand that during electrical stimulation, not all the decrease of force output is 'fatigue'. Under some circumstances (Fig. 2C), the decrease of force output might result from reduced neuron excitability (voltage threshold increased when the efflux of $K^+$ is facilitated by external electrical stimulation). In such case, this decrease of force output could be recovered, and even reversed to an increase of force output when external electrical stimulation is continuously applied.

So far, our investigation is limited to how external electrical stimulation shifts neuron excitability. Since external electrical stimulation shifts neuron excitability by distorting transmembrane ion distribution, action potentials formed by transmembrane ion flows will also shift local neuron excitability. Then, in the following sessions, we build a dynamic model to study the generation and propagation of action potentials in electrical stimulation on the peripheral nerves. This dynamic model helps to elucidate important phenomena, such as collision blocking (25-29).

**A systematic explanation of collision blocking during peripheral nerves electrical stimulation using a bipolar configuration.** Collision blocking has been widely explored to elicit unidirectional propagated action potentials on the peripheral nerves, which is conventionally explained as introducing a stream of antidromic action potentials which annihilate by collision the naturally arising activity (25-29). However, the available explanation of collision blocking does not provide a system-level explanation taking consideration of all the important parameters, including the electrode configuration, input current waveforms. Here, from the standpoint of how external electrical stimulation and the generated action

potentials shifts neuron excitability, we build a system-level model to study the dynamic interaction between the stimulation electrodes and the peripheral nerve. With this dynamic model, phenomena of collision blocking can be easily understood.

**A simplified discussion on positive monophasic stimulation.** Here, we take the bipolar stimulation on the sciatic nerve as an example (Fig. 7A). The bipolar electrodes consist of electrode A (connected to the negative terminal of the stimulator) on the proximal end, and electrode B (connected to the positive terminal of the stimulator) on the distal end of the sciatic nerve. The electroneurogram (ENG) signal on site C is used to indicate the propagation of efferent impulses. Since there are many parameters affecting the generation and propagation of action potentials, we will only study the positive monophasic input current waveform. With the knowledge of how positive monophasic input current affects the generation and propagation of action potentials, other complicated waveforms can then be intuitively understood.

Two possible results are expected to be observed on cite C: with or without action potential. When action potentials can be observed on cite C, these action potentials may result from the successful propagation of either action potentials generated by electrode A ($AP_A$) or action potentials generated by electrode B ($AP_B$), or both $AP_A$ and $AP_B$. In the case of no action potentials generated on cite C, the situation can be two cases. Firstly, the external stimulation may be too week to induce any efferent impulses. Secondly, the external stimulation induces efferent impulses, but these efferent impulses fail to successfully propagate to cite C.

**Fundamental knowledge about bipolar stimulation configuration.** In our earlier static distributed-parameter circuit model (21), bipolar stimulation configuration has been thoroughly studied. There are three important facts about bipolar stimulation configuration, which will serve as the fundamental knowledge to study the generation and propagation of action potentials in this paper.

1. Neurons at each stimulation electrode are independently recruited by that electrode. Therefore, there is an area stimulated by electrode A (red area in Fig. 7A), and another area stimulated by electrode B (yellow area in Fig. 7A).

2. Input current waveforms (and the generated voltage waveforms) at the two stimulation electrodes are of the opposite polarity. In our current study, the voltage waveform at positive electrode B is positive-first, while the voltage waveform at negative electrode A is negative-first.

3. An action potential can be activated, when the extracellular potential becomes negative. Thus, during the input of the positive monophasic current, neurons at electrode A can be activated, while the neurons at electrode B will not be activated. However, due to the resonance caused by RLC components, the extracellular potential at electrode B might be negative enough to activate neurons when the large-amplitude positive input finishes.

With the considerations of these three facts about bipolar stimulation configuration, we have a rough idea about the generation of action potentials at the two electrodes. We will study in detail how the three important parameters (current amplitude, pulse width, electrode distance) affect the generation and propagation of action potentials.

**Three important parameters that affect the generation and propagation of action potentials.** Whether action potentials can be observed on site C is determined by the generation and propagation of action potentials at the two stimulation electrodes, which is affected by three important parameters, namely, current amplitude, pulse width, and electrode distance. Here, we will study in detail how these parameters individually contribute to the generation and propagation of action potentials.

1. Current amplitude.

Current amplitude affects the generation of action potentials at electrode A in two ways. Firstly, higher current amplitude generates an earlier action potential. This is because higher current amplitude induces higher voltage amplitude that exceeds voltage threshold to a larger extent, so that it induces an earlier action potential. Secondly, higher current amplitude generates a frontier action potential A (frontier $AP_A$) that is closer to electrode B (Fig. 7B1, Fig. 7B2). We study this frontier $AP_A$, because it is the only action potential to successfully propagate towards electrode B. This is because all the other action potentials generated at electrode A will annihilate, due to the refractory period of ion channels on their way of propagation. Thus, considering these two effects, higher current amplitude generates an earlier frontier $AP_A$ that is closer to electrode B, so that this frontier $AP_A$ can arrive at electrode B earlier. If this frontier $AP_A$ arrives at electrode B before the positive external stimulation finishes, it will directly interact with this external electrical stimulation.

However, instead of direct interaction with the external stimulation, the frontier $AP_A$ may also interact with the action potentials generated at electrode B. Although the positive input current delivered from electrode B cannot depolarize neurons, the resonance when the positive input current finishes may induce voltage waveform that is negative enough to generation action potentials (20, 21). Whether the negative voltage waveform caused by resonance is strong enough to generate action potentials is largely determined by the current amplitude. Larger current amplitude generates larger resonance. In this way, the generated action potentials at electrode B can also interact with the frontier $AP_A$, to either facilitate or hinder the propagation of this frontier $AP_A$ (depending on the phase of $AP_B$ when the interaction with the frontier $AP_A$ starts).

2. Pulse width.

Pulse width also affects the generation of action potentials at electrode A. As both the current amplitude and pulse width affects the probability calculation, an extreme case is that the action potentials won't be generated even at very high current amplitude when the pulse width is too short.

In addition, pulse width affects how the frontier $AP_A$ will interact with electrode B. The first case is that the frontier $AP_A$ won't be affected by electrode B. If pulse width is too short, when the frontier $AP_A$ arrives at electrode B, it will just successfully propagate towards the observation site C. The second case is the frontier $AP_A$ will interact with external electrical stimulation delivered from electrode B. This will happen when pulse width is long, so that external electrical stimulation is still delivered from electrode B when the frontier $AP_A$ arrives. The third case is the frontier $AP_A$ will interact with the generated $AP_B$. This will happen when the pulse width is of a proper value in between the above two cases.

3. Electrode distance.

The distance between the two stimulation electrodes affects how long the frontier $AP_A$ needs to travel to start interaction with electrode B. This distance together with the generation time of the frontier $AP_A$ (affected by current amplitude) and the pulse width, will determine how the frontier $AP_A$ interacts with electrode B. One extreme case is that the two stimulation electrodes are too far away, so that the frontier $AP_A$ will always successfully propagate towards observation site C, without any interaction with electrode B.

Therefore, when we study the interaction of $AP_A$ and electrode B, we need to know both the timing (when $AP_A$ is generated, how fast $AP_A$ propagates, and whether the external stimulation from electrode B is finished) and the neural excitability around electrode B. The neural excitability around electrode B can be affected by both external electrical stimulation and $AP_B$, and it will determine whether the propagation of $AP_A$ is facilitated or hindered. Thus, current amplitude, pulse width, and electrode distance all contribute to the phenomena of collision blocking.

In summary, our contribution in this manuscript is two-fold. Firstly, we confirmed that external electrical stimulation shifts neuron excitability. Secondly, from the standpoint of how external electrical stimulation and the generated action potentials affect neuron excitability, we propose a system-level dynamic model to study the generation and propagation of action potentials during external electrical stimulation. With a thorough understanding of this dynamic model, important phenomena like collision blocking can be intuitively understood.

Our discovery provides conclusive evidence that external electrical stimulation induces disturbance on transmembrane ion distribution. This disturbance is an important reason to account for the unnatural neuron excitation by electrical stimulation, as compared to the voluntarily controlled movements. The confirmation of this disturbance indicates new designs of the next-generation neuromodulation, which aims at mimicking the natural action potentials to diminish the disturbance caused by external electrical stimulation.

**Methods**

***In vivo* electrode implantation.** All experiments were conducted according to protocols approved by the Institutional Animal Care and Use Committee at the National University of Singapore. Sprague-Dawley rats (around 450 g) were used for the acute experiments. Anesthesia (Aerrane®, Baxter Healthcare Corp., USA) was induced with isoflurane. Carprofen (Rimadyl®, Zoetis, Inc., USA) was injected for pain relief before surgery. After the rat was anesthetized, fur on the leg was gently removed by a shaver. Then, the skin was disinfected with 70% ethanol wipes, and an incision was made with a surgical blade to expose the Tibialis Anterior (TA) muscle. Two stainless steel wires were sutured perpendicular to the muscle fiber direction, with a separation distance of around 1 cm and a depth into the muscle of 2-3 mm.

**Electrical stimulation.** Enveloped high frequency stimulation waveform was generated by MATLAB (MathWorks, USA), to control a Data Acquisition device (DAQ, National Instruments, USA). The voltage output generated by the DAQ was connected to an isolated high-power stimulator (A-M SYSTEMS model 4100, USA), to generate current output to stimulate the TA muscle.

**Force data collection and analysis.** The anesthetized rat was fixed on a stand, and the ankle of left leg was connected to a dual-range force sensor (Vernier, USA). This force sensor was connected to a laptop through a DAQ. LabView (National Instruments, USA) was used for on-site result visualization during the measurements. After the measurements, MATLAB was used for data analysis.

**Distributed-parameter circuit modeling.** The modeling was performed on MATLAB. Firstly, a circuit description was performed in Simulink (MathWorks, USA). Then, current inputs of different waveforms were recursively fed to the circuit model, and the voltage responses of the targeted RLC component were collected. Lastly, these voltage responses were fed into the probability equation to calculate the probability of neuron excitation under these current inputs.

**Acknowledgements**

We thank Dr. Shih-Cheng Yen for helpful discussions on the manuscript, and Han Wu, Li Jing Ong , Dr. Wendy Yen Xian Peh for the experiment setup support. We also thank Gammad Gil Gerald Lasam for the animal experiment support. This work was supported by the following grant from the National Research Foundation: Competitive Research Project 'Peripheral Nerve Prostheses: A Paradigm Shift in Restoring Dexterous Limb Function' (NRF-CRP10-2012-01), and the grant from the HIFES Seed Funding: 'Hybrid Integration of Flexible Power Source and Pressure Sensors' (R-263-501-012-133).

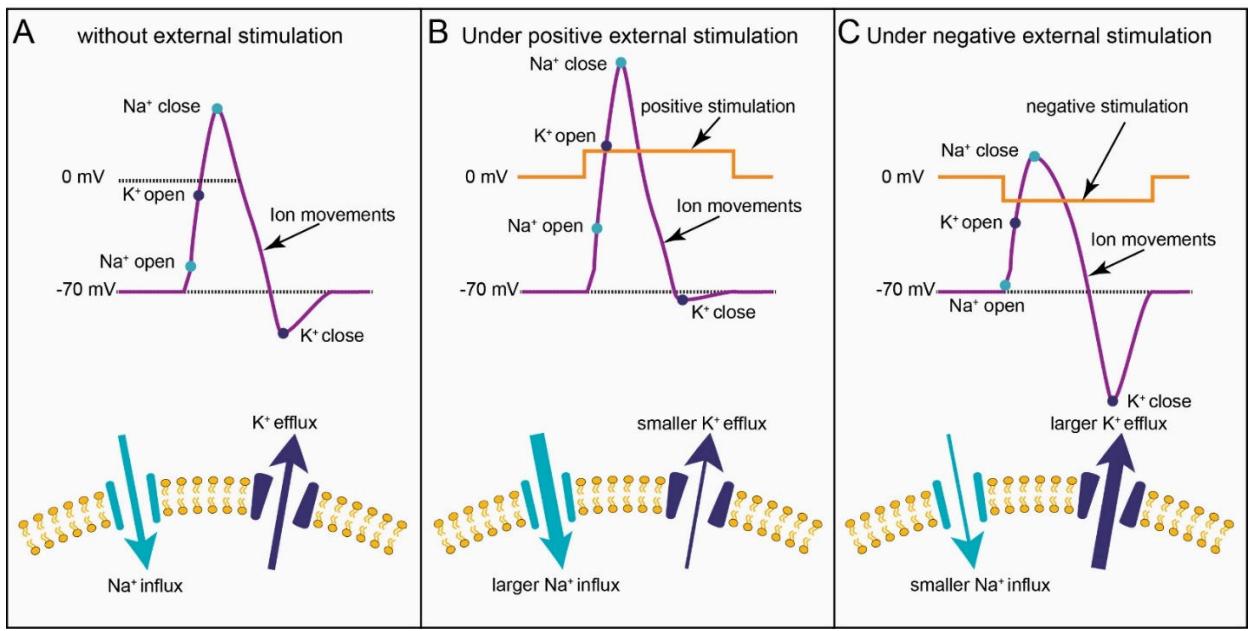

Fig. 1. Illustration of transmembrane ion movements induced by natural action potential and external electrical stimulation. The purple curve shows the transmembrane voltage difference ($V_m = V_{intracellular} - V_{extracellular}$) induced by transmembrane ion movement. (A) Transmembrane ion movement during a natural action potential. (B) Transmembrane ion movement during an action potential with the presence of positive external stimulation. A sufficiently negatively-charged extracellular media requires a larger $Na^+$ influx and a smaller $K^+$ efflux. This reduces the threshold for the following stimulation. (C) Transmembrane ion movement during an action potential with the presence of negative external stimulation. A smaller $Na^+$ influx and a larger $K^+$ efflux are required for a sufficiently negatively-charged extracellular media. This increases the threshold for the following stimulation.

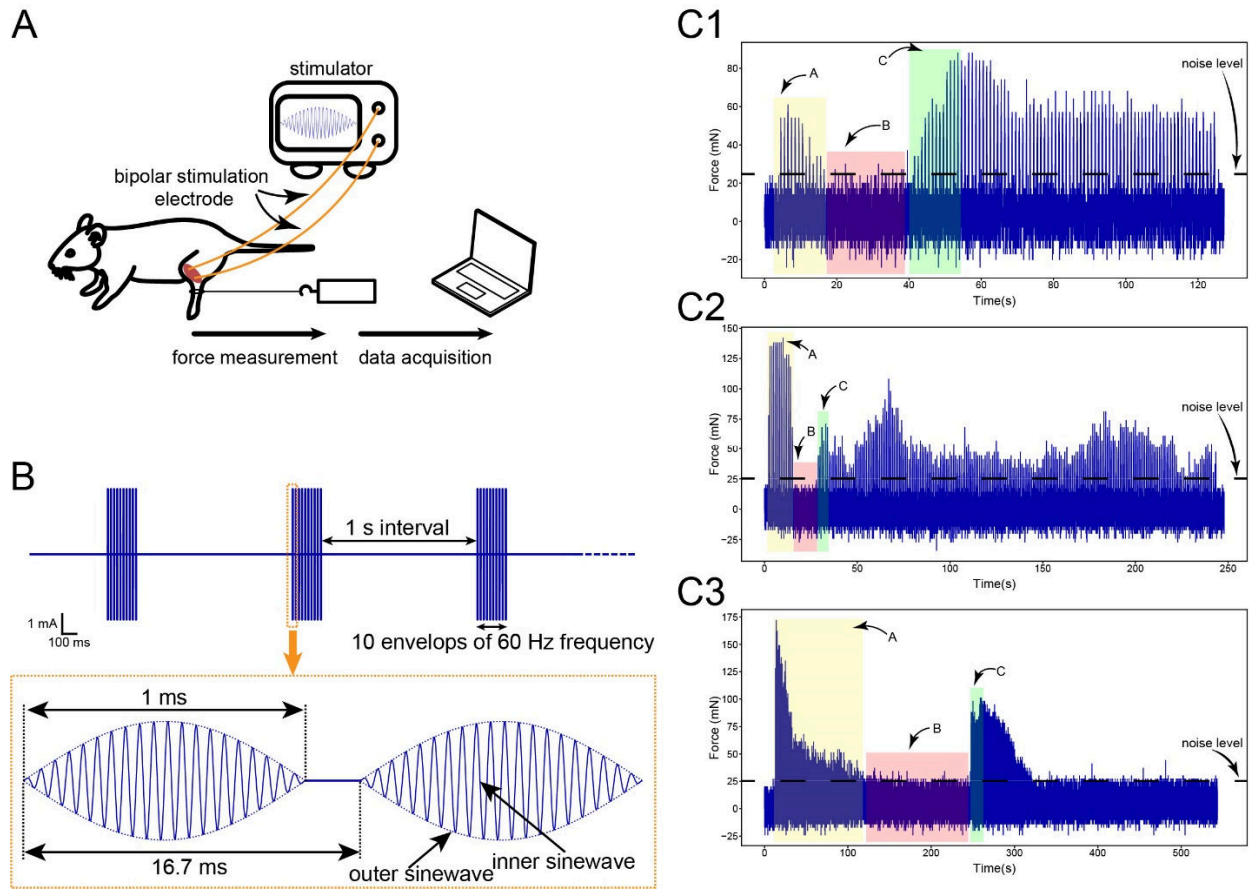

Fig. 2. Measurement setup, Enveloped High Frequency Stimulation (EHFS) waveform, and the generated force profiles. (A) Experiment setup. Two stainless steel wires were implanted in the Tibialis Anterior (TA) muscle for bipolar electrical stimulation. A force gauge was tied to the leg to record force during electrical stimulation. (B) Every 1 s, a train of 10 envelops was delivered. Among the 10 envelops, one envelop was delivered every 16.7 ms. Each envelop consisted of multiple high frequency inner sinewave current pulses (20 kHz). The amplitude of these inner sinewave pulses was modulated by a relatively low frequency outer sinewave (each envelop lasted for 1 ms). (C1-C3) Force profiles generated by EHFS in three *in vivo* measurements show large decrease and increase. Area A refers to the initial force at the beginning of stimulation. Area B refers to the zero force output, which is caused by decreased neuron excitability. Area C refers to the recovery of force output, which is caused by increase neuron excitability. Here, force output under noise level is considered as zero output.

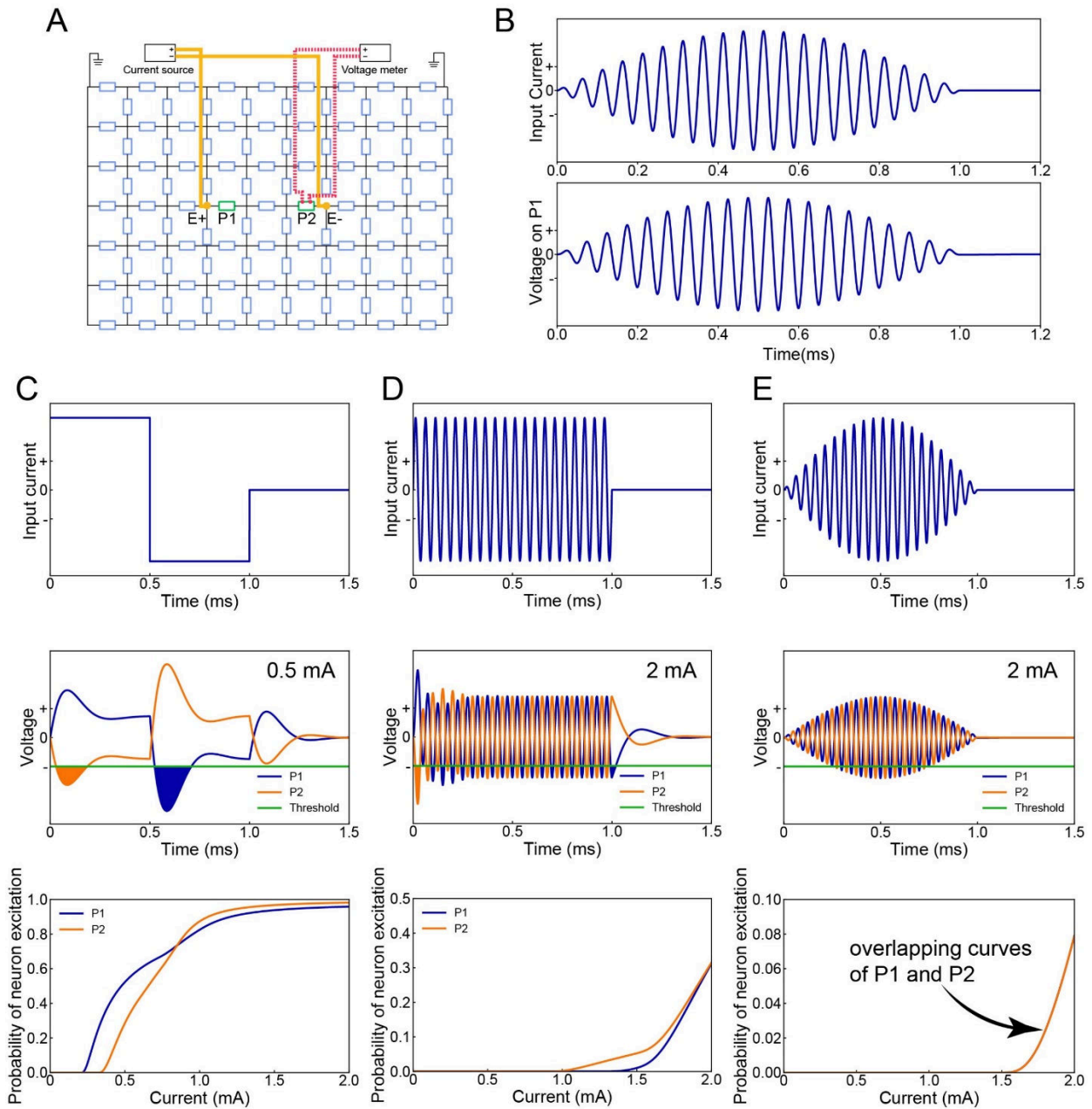

Fig. 3. Voltage waveforms in response to different current input. (A) Distributed-parameter circuit for voltage waveform modeling. E+ and E- are electrode sites to deliver current input. P1 and P2 are two representative motoneurons which are close to E+ and E- respectively. (B) EHFS input current and the voltage waveform measured from P1. (C, D, E) Input current waveform, voltage waveforms and probability curves of P1 and P2 with input current of square wave (C), 20 kHz sinewave (D), and enveloped 20 kHz sinewave (E). For square wave current input, voltage waveforms show different pattern of exceeding the threshold for neurons at positive electrode (P1) and negative electrode (P2). For 20 kHz sinewave, voltage waveforms measured from P1 and P2 only show asynchrony at the beginning and ending of the high frequency pulse train. For enveloped 20 kHz sinewave, voltage waveforms measured from P1 and P2 are well synchronized.

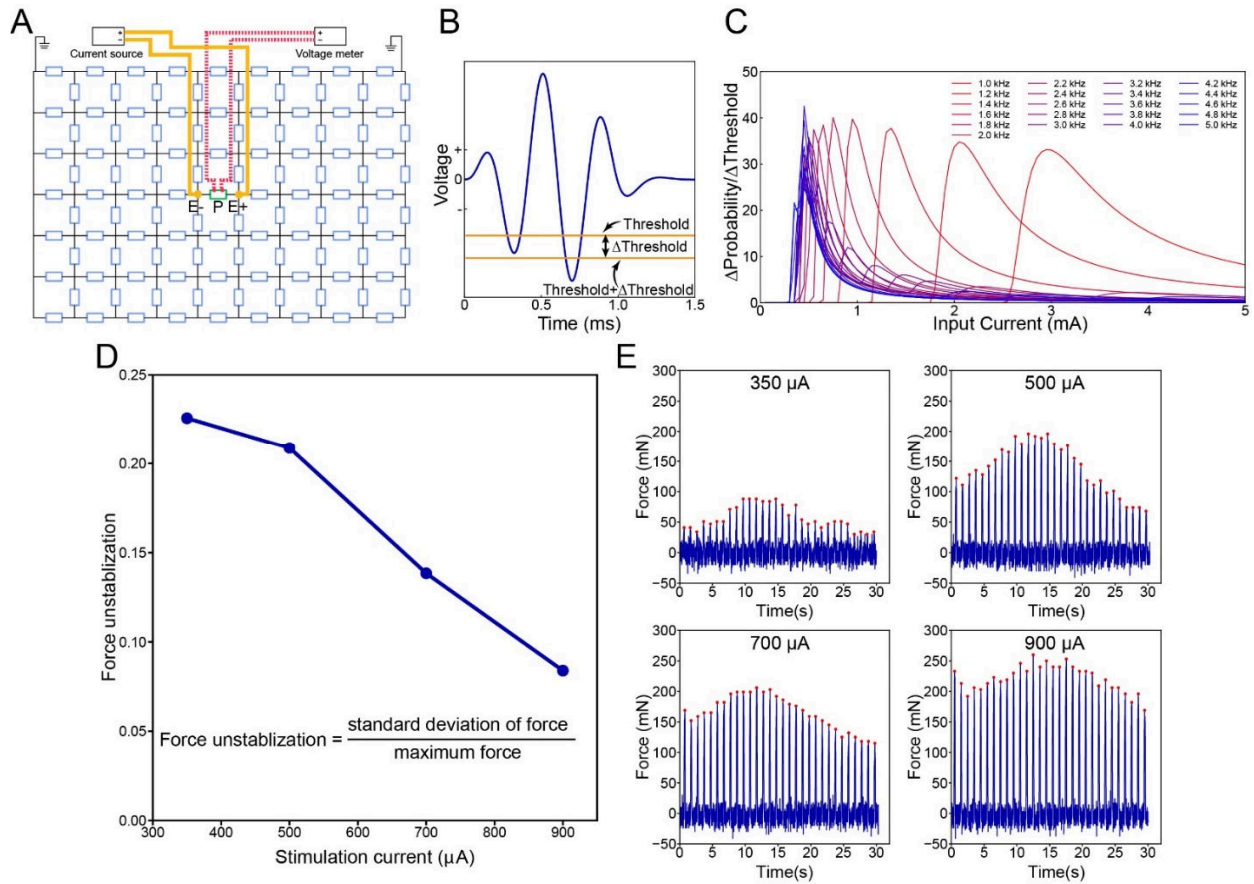

Fig. 4. Modeling and *in vivo* measurement of how current amplitude affects the stability of force output. (A) Distributed-parameter circuit with two stimulation electrodes (E+ and E1) and a motoneuron (P). (B) Illustration of voltage waveform, threshold voltage, and threshold voltage shift. (C) $\Delta Probability = Probability_{Threshold} - Probability_{Threshold+\Delta Threshold}$. At all frequencies of input current, the probability of neuron excitation becomes more resistive to threshold voltage shift at large current amplitude. (D) Quantification of force unstabilization at different current. (E) Force profiles measured at different current amplitude. Large current amplitude generates more stable force output.

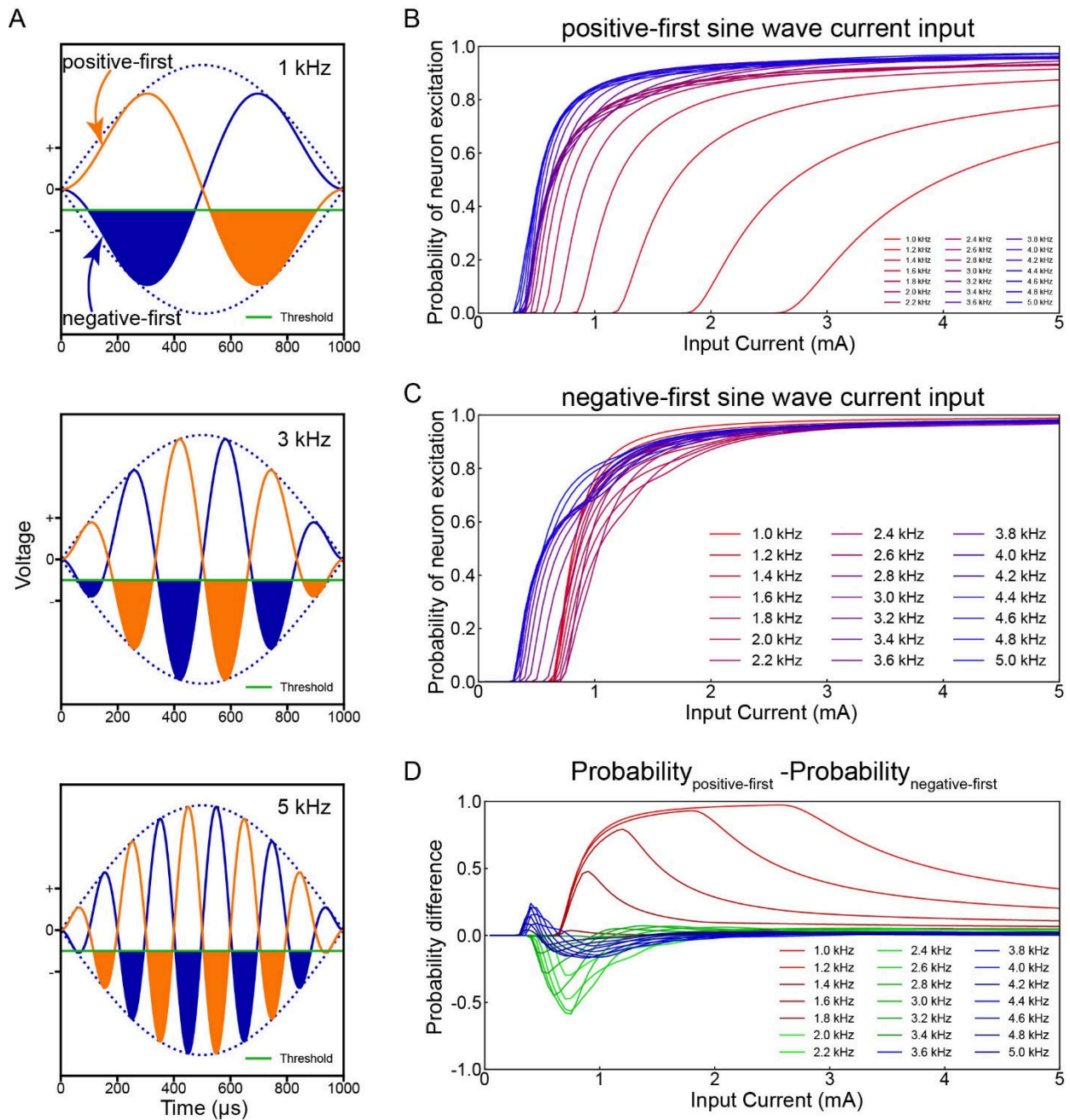

Fig. 5. Modeling of how probability difference in response to positive-first and negative-first current diminishes with increasing current amplitude and frequency. (A) Voltage waveforms in response to EHFS of 1 kHz, 3 kHz, and 5 kHz. (B, C) Probability at different current amplitude and frequency using input current of positive-first EHFS (B), and negative-first EHFS (C). (D) Probability difference diminishes when input current is of high frequency and high amplitude.

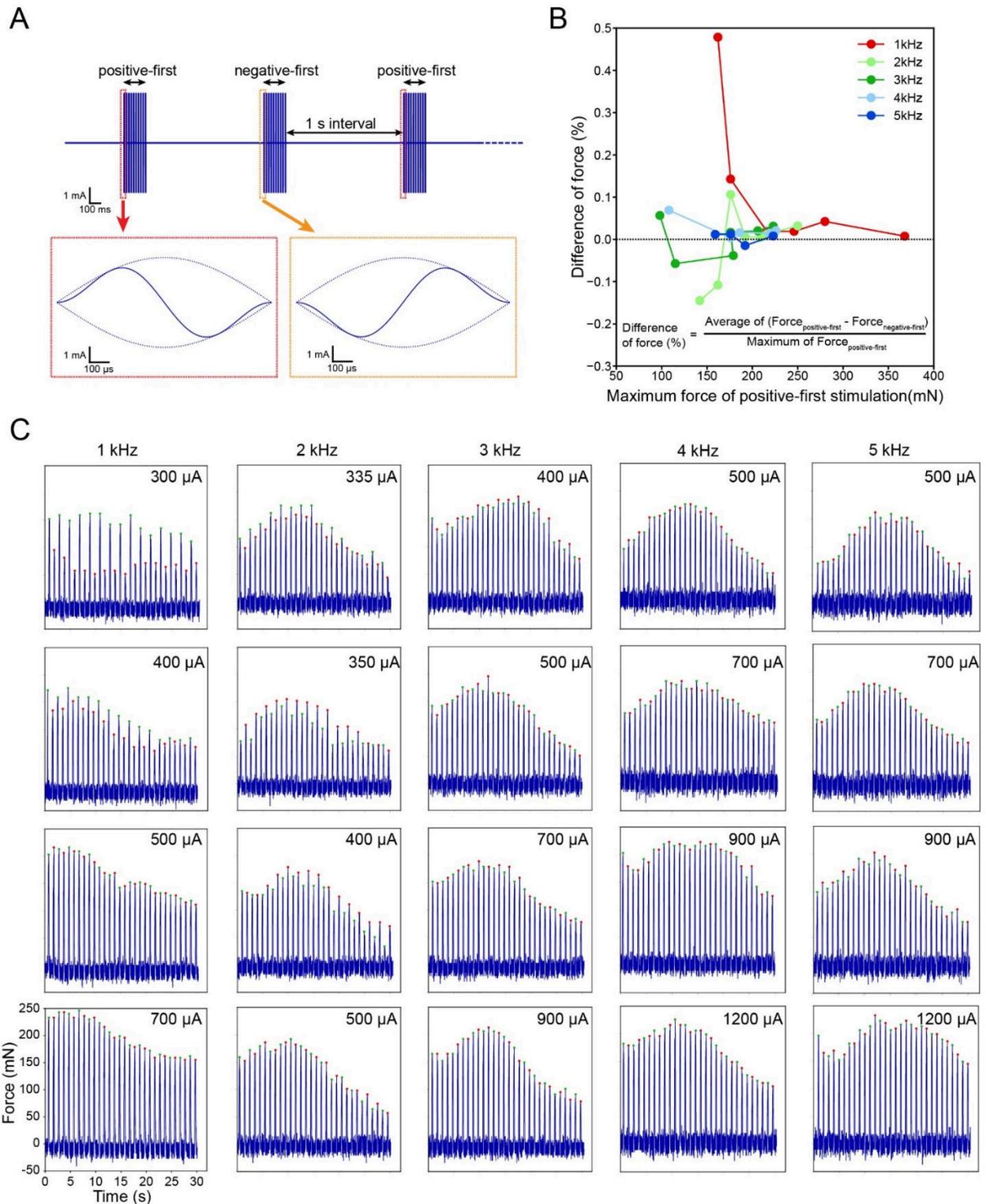

Fig. 6. Motoneuron activation at positive and negative electrode becomes more synchronized with increasing amplitude and frequency of input current. (A) Illustration of input current. Positive-first and negative-first EHFS is alternatively delivered every 1 s. (B) Quantification of the force difference using positive-first and negative-first EHFS. (C) Force profiles measured at input current of different frequency

and amplitude. The peak forces are alternatively marked as green or red point, to indicate the force generated by positive-first and negative-first EHFS.

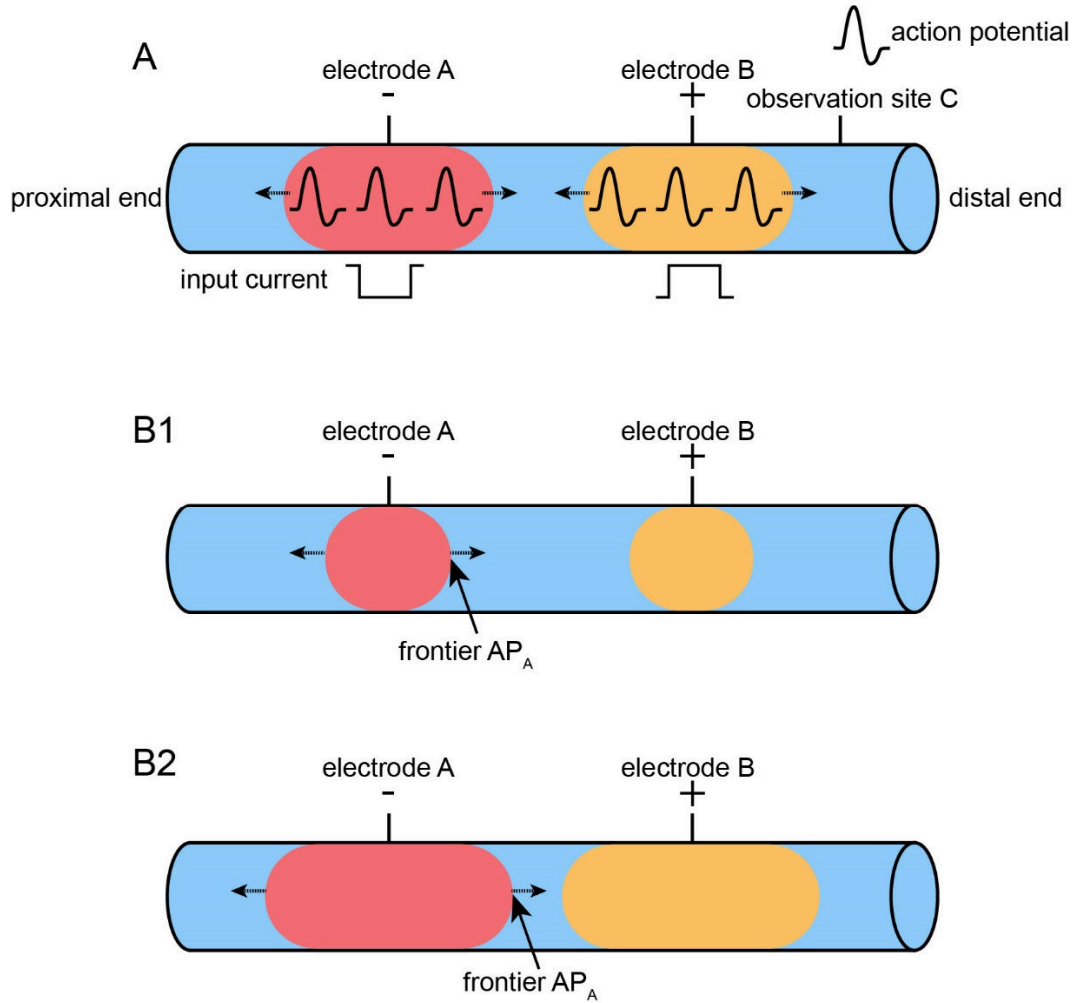

Fig. 7. Illustration of bipolar stimulation on the sciatic nerve. (A) The bipolar configuration consists of electrode A (connected to the negative terminal of the stimulator) on the proximal end of the sciatic nerve, and electrode B (connected to the positive terminal of the stimulator). Positive monophasic input current is delivered to the nerve. Input current at the two stimulation electrodes are of the opposite polarity. The two stimulation electrodes will independently activate motoneurons around them. (B1, B2) The stimulation area around the electrode is smaller at lower current amplitude (B1), and larger at higher current amplitude (B2). In both cases, the frontier $AP_A$ propagates towards electrode B.

Supplementary information

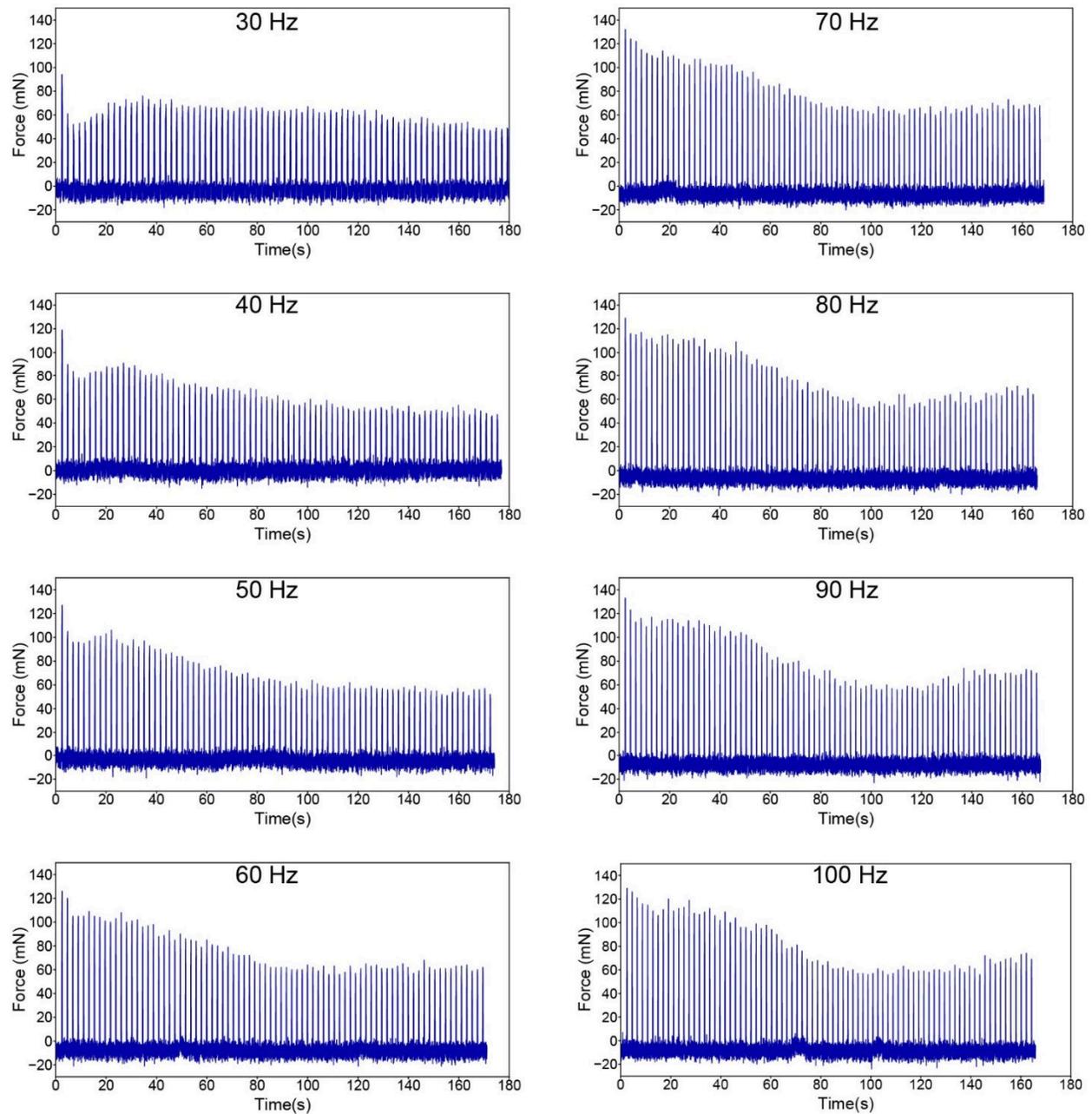

Fig. S1. Force profiles with fluctuations generated by continuous square wave stimulation. Every 1 s, 10 square wave current pulses were delivered at 30-100 Hz (an envelope was delivered every 33.3-10 ms). These force profiles measured with square wave current input also show fluctuation, but not as significant as induced by enveloped high frequency stimulation.

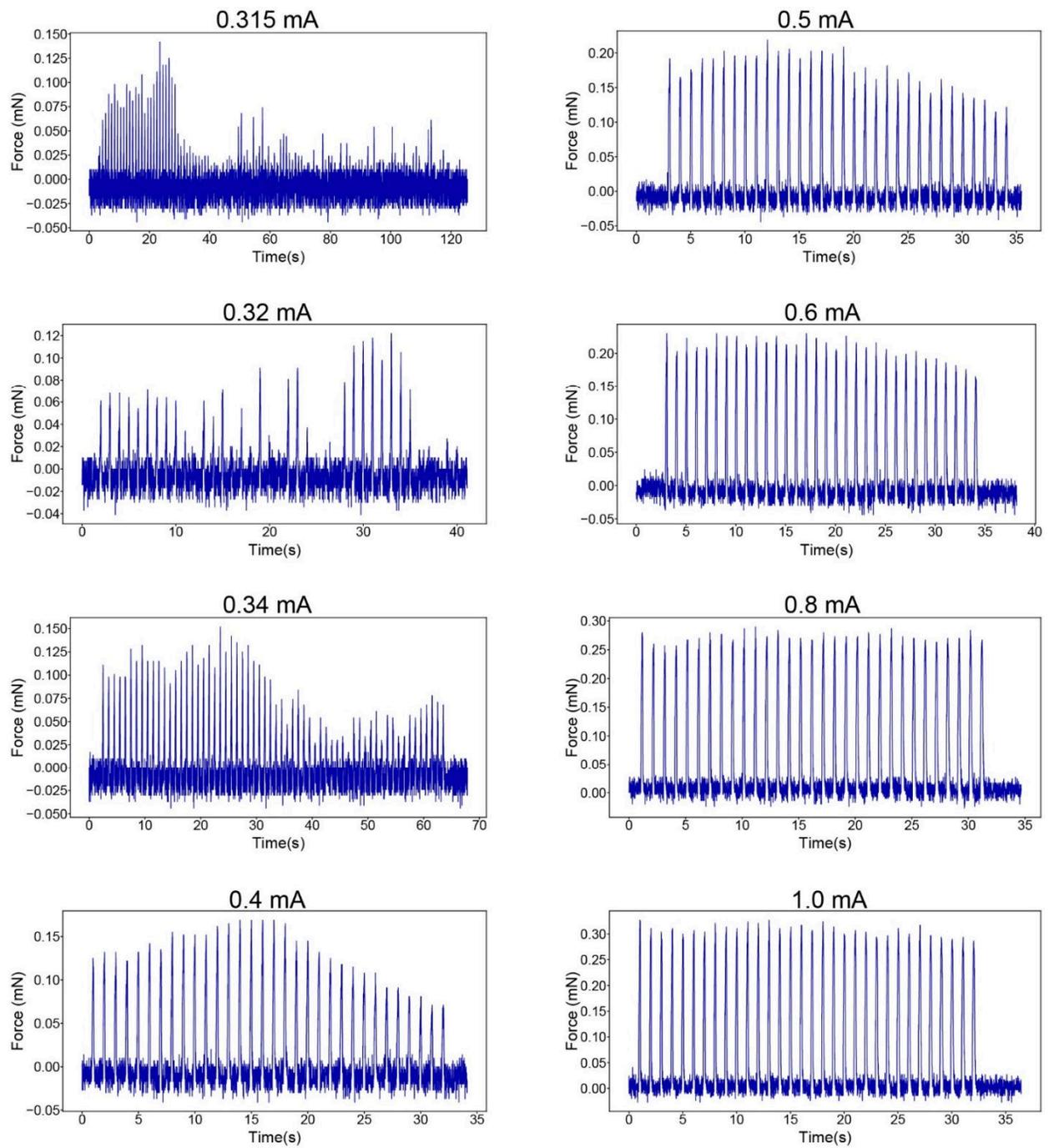

Fig. S2. Force profiles generated with enveloped high frequency sinewave pulses (3 kHz) at different current amplitudes, measured from the same object. It is clear that low input current amplitude generates unstable force output of low magnitude, while high input current amplitude generates stable force output of high magnitude.

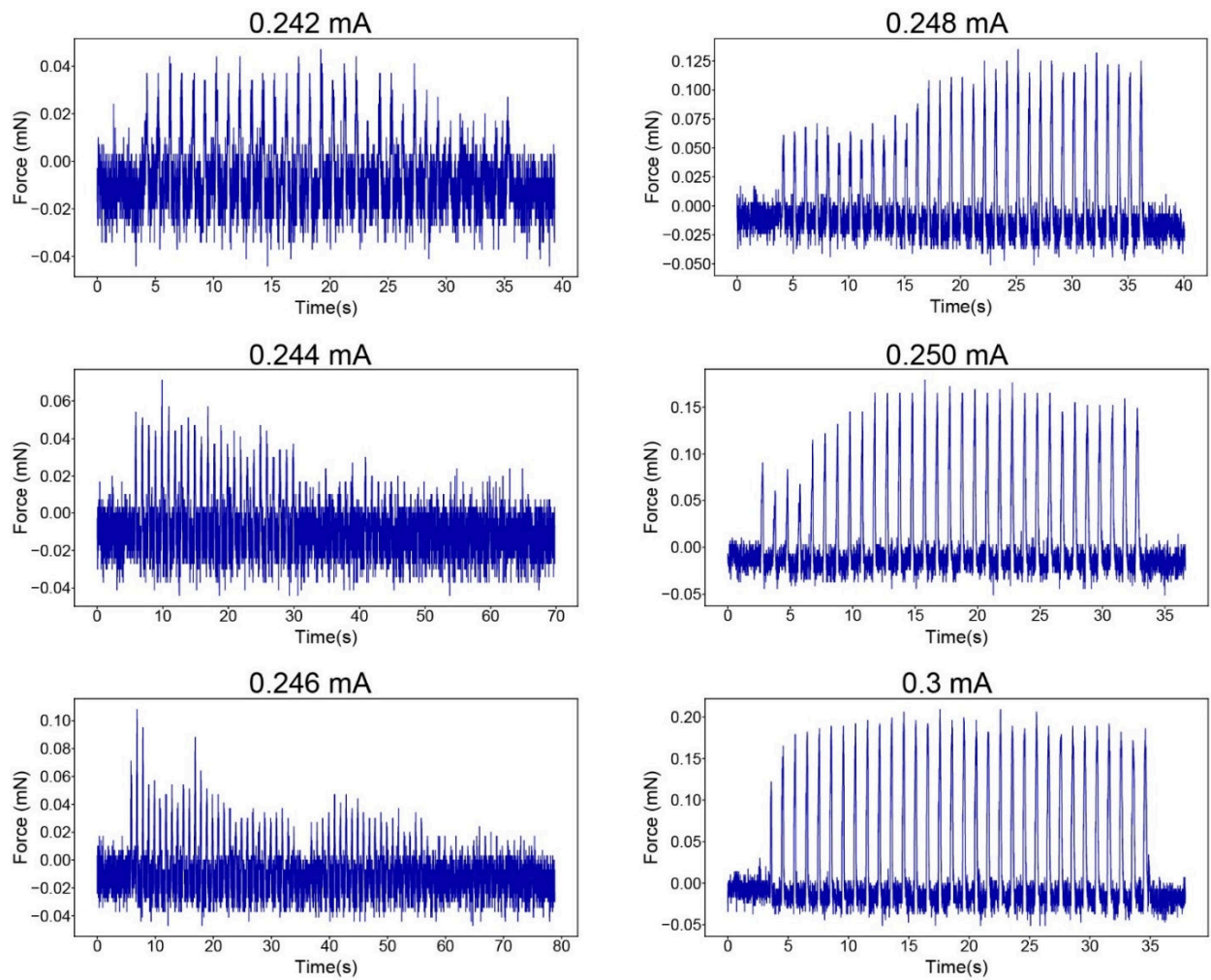

Fig. S3. Force profiles generated with enveloped high frequency sinewave pulses (3 kHz) at different current amplitudes, measured from another object. It is also clear that low input current amplitude generates unstable force output of low magnitude, while high input current amplitude generates stable force output of high magnitude.

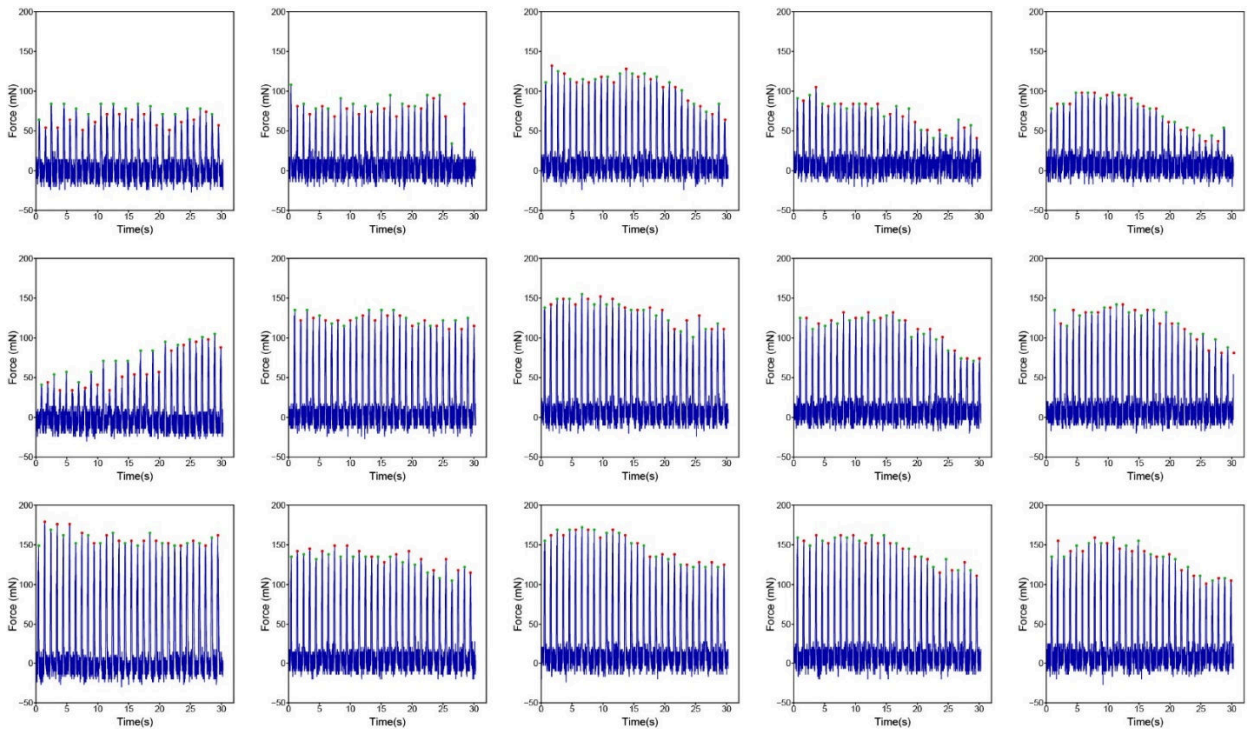

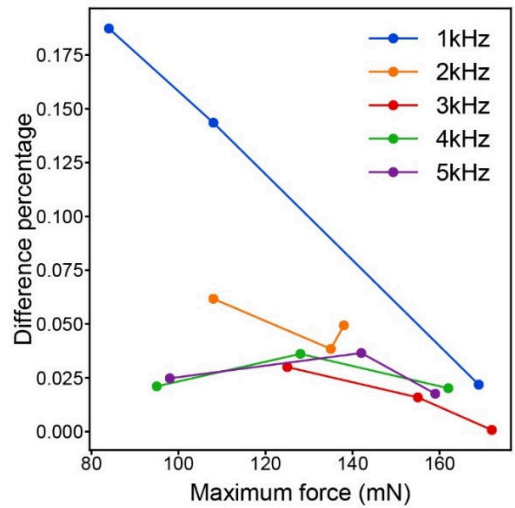

Fig. S4. Asymmetrical force output using positive-first and negative-first enveloped high frequency stimulation, measured from another object. These measurement results are in agreement with Fig. 5 in two ways. First, the magnitude of force asymmetry decreases with increasing frequency. Second, the force asymmetry decreases with increasing current amplitude. However, these measurement results fail to show the polarity shift as shown in Fig. 6. This could be explained if the frequency range in this object that could enable the observation of polarity shift is small, and we failed to catch a frequency than fall in this specific range.